\documentclass[english,reprint,amsmath,amssymb,aps,prb]{revtex4-2}
\usepackage[T1]{fontenc}
\usepackage[utf8]{inputenc}

\usepackage{graphicx}
\usepackage[english]{babel}
\usepackage{xcolor}
\usepackage[separate-uncertainty=true]{siunitx}
\usepackage{physics}
\usepackage[version=4]{mhchem}

\usepackage[colorlinks=true,citecolor={blue!70!black},linkcolor=black,urlcolor=black]{hyperref}
\usepackage{xr}

\begin{document}

\title{Fermion parity of an Andreev molecule probed by nonlocal Josephson effect}

\author{S. Annabi$^{1}$}
\author{H. Riechert$^{1}$}
\author{K. Watanabe$^{2}$}
\author{T. Taniguchi$^{3}$}
\author{J. Griesmar$^{1}$}
\author{E. Arrighi$^{4}$}
\author{L. Bretheau$^{1*}$}
\author{J.-D. Pillet$^{1}$}

\altaffiliation{These authors jointly supervised this work.
\newline
jean-damien.pillet@polytechnique.edu
\newline
landry.bretheau@polytechnique.edu}

\affiliation{$^{1}$ Laboratoire de Physique de la Mati\`ere condens\'ee, CNRS, École polytechnique, Institut Polytechnique de Paris, 91120 Palaiseau, France}
\affiliation{$^{2}$ Research Center for Electronic and Optical Materials, National Institute for Materials Science, 1-1 Namiki, Tsukuba 305-0044, Japan}
\affiliation{$^{3}$ Research Center for Materials Nanoarchitectonics, National Institute for Materials Science,  1-1 Namiki, Tsukuba 305-0044, Japan}
\affiliation{$^{4}$ Brazilian Nanotechnology National Laboratory (LNNano), Brazilian Center for Research in Energy and Materials (CNPEM), Campinas, São Paulo 13083-970, Brazil}

\begin{abstract}
Fermion parity is a fundamental property of superconducting many-body states. Here, we show that the global fermion parity of a delocalized superconducting state can be detected locally by exploiting the nonlocal Josephson effect. Using a carbon nanotube-based Andreev molecule formed by two coupled quantum-dot Josephson junctions, we observe a pronounced nonlocal Josephson response and demonstrate the formation of delocalized Andreev molecular states extending across both junctions. We further show that changes in the molecular ground-state parity manifest as characteristic $\pi$-phase shifts in the nonlocal response. Supported by a minimal theoretical model, these results identify global fermion parity as an experimentally accessible degree of freedom in hybrid superconducting circuits that can be readily revealed through the nonlocal Josephson effect.
\end{abstract}

\maketitle

Fermion parity is a fundamental quantum number in superconductivity, as it determines the microscopic nature of the many-body ground state~\cite{matveev_effects_1994, joyez_observation_1994, matveev_parity_1997, catelani_decoherence_2012, catelani_parity_2014, gustavsson_suppressing_2016, krause_quasiparticle_2024}. In a Josephson junction, this parity is encoded in the occupation of Andreev bound states and therefore directly governs the magnitude and direction of the supercurrent~\cite{zgirski_evidence_2011, bretheau_exciting_2013,bretheau_supercurrent_2013, janvier_coherent_2015}. When the weak link is reduced to an interacting quantum dot, fermion parity can be controlled electrostatically through gate voltages~\cite{pillet_andreev_2010, martin-rodero_josephson_2011, lee_spin-resolved_2014, fatemi_microwave_2022, keliri_driven_2023, coraiola_spin-degeneracy_2024, sahu_ground-state_2024}. This gives rise to one of the most emblematic signatures of parity in superconducting systems: the $0$--$\pi$ transition, corresponding to a reversal of the Josephson current~\cite{van_dam_supercurrent_2006,jorgensen_critical_2007,delagrange_0-ensuremathpi_2016}. Microscopically, this effect originates from a quantum phase transition driven by the crossing of Andreev states with different parity, resulting in an abrupt change of the many-body ground state~\cite{vecino_josephson_2003, bauer_spectral_2007, meng_self-consistent_2009, deacon_tunneling_2010, maurand_first-order_2012, pillet_tunneling_2013, su_andreev_2017, estrada_saldana_supercurrent_2018, bouman_triplet-blockaded_2020}. While parity is inherently a local quantity in a single Josephson junction, the situation becomes fundamentally different in more complex superconducting networks as Andreev states can hybridize and delocalize across several coupled junctions~\cite{leijnse_parity_2012, melin_dc_2014, feinberg_quartets_2015, ten_haaf_two-site_2024, johannsen_fermionic_2025}. In this regime, global fermion parity emerges as a natural quantum number, raising the question of how it manifests itself in the Josephson response. Beyond this fundamental interest, controlling and reading out global fermion parity is becoming increasingly relevant for envisioned devices that encode quantum information in delocalized fermionic degrees of freedom~\cite{dvir_realization_2023, ten_haaf_two-site_2024, van_driel_charge_2024, van_loo_single-shot_2026}.

The most elementary system that realizes such a network is the Andreev molecule formed in a pair of coupled Josephson junctions~\cite{pillet_nonlocal_2019, kornich_fine_2019, pillet_josephson_2023}. When these two junctions are brought sufficiently close, their respective Andreev bound states overlap and hybridize into delocalized molecular states. A remarkable manifestation of this hybridization is the nonlocal Josephson effect, whereby the supercurrent flowing through one junction depends periodically on the superconducting phase difference applied across the other. Although the nonlocal Josephson effect has been observed in several hybrid semiconductor-superconductor circuits~\cite{matsuo_observation_2022, matsuo_phase-dependent_2023, haxell_demonstration_2023}, establishing a direct connection between the measured nonlocal response and the spatial delocalization of the underlying quantum state remains challenging. Even more elusive is the role played by the parity of the corresponding many-body state. Most experimental implementations rely on extended weak links supporting multiple conduction channels, introducing additional microscopic degrees of freedom that can obscure the relation between nonlocal transport signatures and the existence of a well-defined Andreev molecular state. A minimal and highly controllable platform is therefore required to unambiguously identify the formation of a delocalized fermionic state and to investigate how its parity manifests itself in the Josephson current.

In this work, we realize an Andreev molecule in its minimal form using a single carbon nanotube hosting two strongly interacting quantum-dot Josephson junctions coupled through a common superconducting electrode. Owing to the ballistic and quasi-one-dimensional nature of the nanotube, this architecture provides a highly controllable platform in which Andreev states can hybridize into a well-defined molecular state. We observe a pronounced nonlocal Josephson effect whose magnitude is maximized when the molecular wavefunction becomes delocalized across both junctions. Furthermore, we show that changes in the molecular ground-state parity are encoded in the phase of the nonlocal response through characteristic $\pi$-phase shifts. This behavior echoes the $0$--$\pi$ transition of a single Josephson quantum dot~\cite{delagrange_0-ensuremathpi_2016}, where parity changes reverse the local supercurrent, but appears here in the phase of a nonlocal Josephson signal. Together, these results both establish the nonlocal Josephson effect as a direct probe of the delocalized Andreev molecular states and demonstrate that global fermionic parity can be inferred locally from transport through a single junction.

\section{Realization of an Andreev molecule}

\begin{figure}[h]
    \includegraphics[page=1,width=\columnwidth]{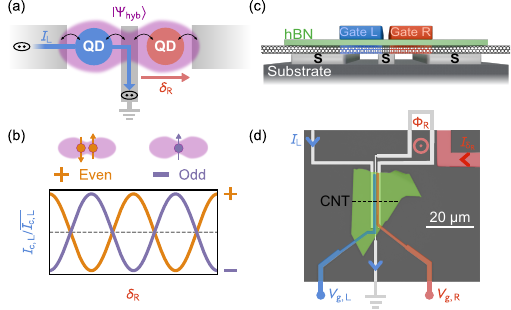}
    \caption{{\bf Fermion parity detection in a carbon nanotube Andreev molecule.} (a)~Schematic of an Andreev molecule formed by two quantum-dot Josephson junctions (blue and red) coupled through a common superconducting electrode (grey). The hybridized molecular state~$\ket{\Psi_\mathrm{hyb}}$ (purple) extends across both junctions and mediates a nonlocal Josephson effect (NLJE), whereby the supercurrent~$I_\mathrm{L}$ through the left junction depends on the superconducting phase difference~$\delta_\mathrm{R}$ across the right junction. (b)~Schematic of parity detection through the NLJE, which manifests as the critical current~$I_\mathrm{c,L}$ of the left junction exhibiting a periodic dependence on~$\delta_\mathrm{R}$. Different parity configurations of the molecular ground state lead to nonlocal responses that are {$\pi$-shifted}. The critical current is normalized by its average value~$\overline{I_\mathrm{c,L}}$ over~$\delta_\mathrm{R}$. (c)~Schematic of the carbon nanotube device. The nanotube lies between a hexagonal boron nitride (hBN) flake and three superconducting electrodes (S). Two local gates (blue and red) provide independent electrostatic control of the two corresponding quantum dots.(d)~False-colored optical image of device~T using the same color code as in~(c). The \SI{100}{\nano\meter}-wide central electrode is grounded while transport is measured through the left junction. The right junction is embedded in a superconducting loop, allowing the phase difference~$\delta_\mathrm{R}$ to be controlled through the magnetic flux~$\Phi_\mathrm{R}$ generated by an on-chip flux line through which a current~$I_\mathrm{\delta_R}$ is applied.
}
    \label{fig1}
\end{figure}

The principle of the experiment is illustrated in Fig.~\ref{fig1}. The system consists of two quantum-dot Josephson junctions connected in series and sharing a common superconducting electrode (Fig.~\ref{fig1}a). When the separation between the two junctions is shorter than the superconducting coherence length, the Andreev bound states of the individual junctions overlap and hybridize into bonding and antibonding molecular states. The resulting many-body ground state is delocalized across both weak links, such that the Josephson response of one junction becomes sensitive to the superconducting phase difference applied across the other. This gives rise to the nonlocal Josephson effect (NLJE), experimentally detected here as a periodic modulation of the critical current of the left junction~$I_\mathrm{c,L}$ with the phase difference $\delta_\mathrm{R}$ imposed across the right one~\cite{pillet_josephson_2023}. Within this picture, changes in the parity of the molecular ground state can lead to a $\pi$-phase shift of $I_\mathrm{c,L}(\delta_\mathrm{R})$, providing a signature of global fermion parity (Fig.~\ref{fig1}b).

To realize such a minimal Andreev molecule, we use an individual carbon nanotube~\cite{annabi_josephson_2024, riechert_carbon_2025} as a common quantum conductor connecting three superconducting electrodes made of a niobium/gold bilayer (Fig.~\ref{fig1}c). This material combines ballistic transport (Supplementary~\ref*{SM:Ballistic}), one-dimensional confinement, and efficient electrostatic tunability, making it an ideal platform for quantum-dot Josephson junctions. The nanotube remains continuous across both junctions, and the central superconducting electrode is sufficiently narrow to place the device in a regime of strong inter-junction hybridization.

The nonlocal response is probed by grounding the central electrode and measuring transport through the left junction while controlling the superconducting phase difference~$\delta_\mathrm{R}$ across the right one. To this end, the right junction is embedded in a superconducting loop~\cite{matsuo_observation_2022, haxell_demonstration_2023}, allowing $\delta_\mathrm{R}$ to be tuned by threading a magnetic flux via an on-chip flux line through which a current~$I_\mathrm{\delta_R}$ is applied (Fig.~\ref{fig1}d). Measurements are performed at \SI{10}{\milli\kelvin} in a dilution refrigerator. The devices are fabricated on a doped silicon substrate serving as a global back-gate electrode that tunes the chemical potential of both quantum dots simultaneously.

To investigate both the parity and spatial structure of the molecular ground state, we study two complementary devices. Device~B is controlled solely by the global back-gate electrode and exhibits a well-developed supercurrent branch, enabling direct measurements of the switching current and its nonlocal modulation. Device~T incorporates two additional local top gates coupled respectively to the left and right junctions, providing independent control of the two quantum dots. In this device, the Josephson response is detected through its zero-bias conductance~\cite{jorgensen_critical_2007} (see Supplementary~\ref*{SM:Conductance_Josephson}).

\section{Nonlocal Josephson response}

\begin{figure}[h]
    \includegraphics[page=1,width=\columnwidth]{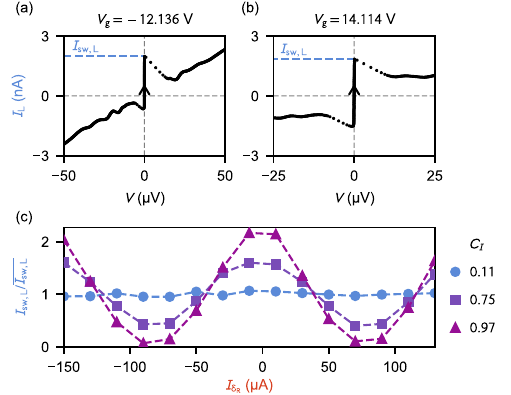}
    \caption{{\bf Nonlocal Josephson effect.} (a)-(b)~Current-voltage characteristics $I_\mathrm{L}(V)$ measured at different back-gate voltages $V_\mathrm{g}$. The current is swept from negative to positive values, and the switching current $I_\mathrm{sw,L}$ is defined as the maximum current sustained by the supercurrent branch. (c)~$I_\mathrm{sw,L}$ as a function of the flux current $I_{\delta_\mathrm{R}}$ controlling the phase difference~$\delta_\mathrm{R}$ across the right junction, shown at three different~$V_\mathrm{g}$ values (\SIlist{14.395;13.938;14.281}{\volt} for respectively circle, square and triangle markers). The periodic modulation demonstrates a pronounced nonlocal Josephson response that is quantified with a modulation contrasts~$C_I$ extracted for each curve.}
    \label{fig2}
\end{figure}

We first establish the existence of a strong nonlocal Josephson response in device~B, which forms the basis of the parity measurements presented later. In this device, a single back-gate electrode controls both quantum dots simultaneously. As shown in Fig.~\ref{fig2}a-b, superconducting transport manifests as a well-developed supercurrent branch in the current-voltage characteristic, whose amplitude can be continuously tuned with gate voltage, as expected for a quantum-dot Josephson junction~\cite{jarillo-herrero_quantum_2006,cleuziou_carbon_2006,annabi_josephson_2024} (see Supplementary~\ref*{SM:Isw_CI_vs_gate}). The switching current $I_\mathrm{sw,L}$ is defined as the maximum current of this nondissipative branch.

The hallmark of the Andreev molecular state appears when the superconducting phase difference across the right junction is varied~\cite{pillet_nonlocal_2019, matsuo_observation_2022, haxell_demonstration_2023}. As shown in Fig.~\ref{fig2}c, the switching current of the left junction exhibits a pronounced periodic modulation as a function of the current~$I_\mathrm{\delta_R}$ applied to the flux line controlling the phase difference~$\delta_\mathrm{R}$ across the right junction (see also Supplementary~\ref*{SM:Switching_measurement}). The observed period corresponds to one superconducting flux quantum ${\Phi_0=h/2e}$ threading the loop (see Methods), demonstrating that the Josephson response of the left junction is directly influenced by the phase bias imposed across the right one. This distinctive behavior constitutes a clear signature of the nonlocal Josephson effect in device~B.

To quantify the strength of the nonlocal response, we define the modulation contrast as
\begin{equation}
\label{eqn:contrast}
C_I =
\frac{\displaystyle\max_{\delta_\mathrm{R}} I_\mathrm{sw,L}
- \min_{\delta_\mathrm{R}} I_\mathrm{sw,L}}
{\displaystyle\max_{\delta_\mathrm{R}} I_\mathrm{sw,L}
+ \min_{\delta_\mathrm{R}} I_\mathrm{sw,L}}\cdot
\end{equation}
This contrast depends strongly on gate voltage (see also Supplementary~\ref*{SM:Isw_CI_vs_gate}), ranging from nearly zero to values approaching unity. In the latter regime, the supercurrent through the left junction can be almost suppressed by changing only the superconducting phase difference across the right one. Such a large nonlocal response is remarkable given the minimal nature of the system, which consists of two coupled quantum-dot Josephson junctions hosted within a single carbon nanotube. It demonstrates that strong nonlocal Josephson coupling can emerge in the most elementary realizations of an Andreev molecule.

Although these measurements establish a pronounced nonlocal Josephson response, they do not by themselves reveal its microscopic origin. In particular, the use of a global gate electrode prevents independent control of the two quantum dots and therefore does not allow hybridization to be directly controlled. To determine whether the observed nonlocal response is indeed related to the formation of a delocalized ground state, we now turn to device~T, where local electrostatic control is implemented.

\section{Molecular hybridization}

A direct connection between the nonlocal Josephson effect and the spatial structure of the molecular ground state can be established by controlling the local microscopic properties of the system. To this end, device~T provides independent control over the energy levels of the left and right quantum dots, allowing to map directly the hybridization of the Andreev molecular state.

\begin{figure}[b]
    \includegraphics[width=\columnwidth]{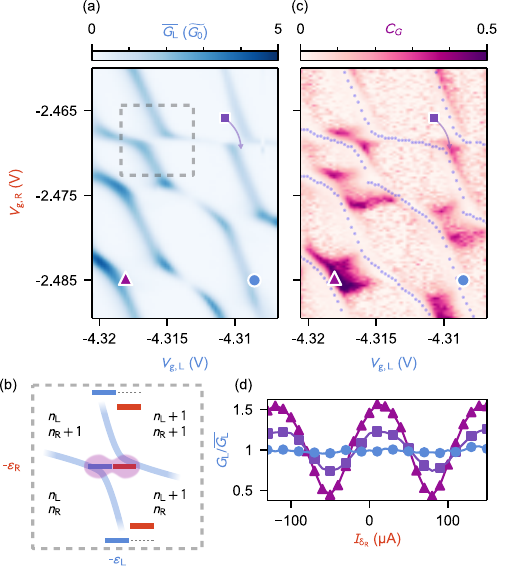}
    \caption{{\bf Hybridization of Andreev molecular states.} (a)~Zero-bias conductance~$\overline{G_\mathrm{L}}$ of the left junction in device~T averaged over the flux current~$I_{\delta_\mathrm{R}}$ as a function of the local gate voltages $V_\mathrm{g,L/R}$. Resonance lines correspond to the alignment of discrete quantum-dot levels with the chemical potential of the superconducting electrodes, providing access to the electronic spectrum of the double quantum dot. Conductance is expressed in units of ${\widetilde{G_0}=10^{-2}e^2/h}$. (b)~Schematic of the region highlighted by the dashed square in (a), where the energy levels $\epsilon_\mathrm{L}$ and $\epsilon_\mathrm{R}$ are tuned into resonance. At the avoided crossing, the two levels hybridize into bonding and antibonding Andreev molecular states delocalized across both junctions (central inset). The charge configurations in the double quantum dot are separated by the blue resonant lines and indicated by the set of integers~$(n_\mathrm{L},n_\mathrm{R})$. (c)~Conductance contrast $C_G$, defined from the normalized amplitude of the flux-dependent modulation shown in (d), plotted as a function of $V_\mathrm{g,L/R}$. Dashed blue lines indicate the resonance lines identified in (a). The strongest nonlocal response occurs at the avoided crossings, where the molecular wavefunction is expected to be maximally delocalized. (d)~Flux-dependent modulation measured at three representative operating points indicated in (c): far from an avoided crossing (circle), at a small avoided crossing (square), and at a large one (triangle). The modulation amplitude increases strongly as the operating point approaches an avoided crossing, linking the strength of the nonlocal Josephson effect to the delocalization of the Andreev molecular state.}
    \label{fig3}
\end{figure}

Here, the Josephson response is detected through its zero-bias conductance signal (Supplementary~\ref*{SM:Conductance_Josephson}). Figure~\ref{fig3}a shows the resulting conductance map as a function of the two local gate voltages. This measurement exhibits the characteristic pattern of a stability diagram, thus providing a direct access to the electronic spectrum of the double quantum dot~\cite{kocsis_strong_2024, kurtossy_heteroatomic_2026}. Each resonance line corresponds to the alignment of a discrete quantum-dot level with the chemical potential of the superconducting electrodes (see also Supplementary~\ref*{SM:DQD_stability_wide}). Two families of resonances can be distinguished, associated with the left and right quantum dots, whose intersections develop into avoided crossings. Such avoided crossing is illustrated schematically in Fig.~\ref{fig3}b. At the center of the anticrossing, the energy levels of the two quantum dots become resonant and hybridize. In the presence of superconducting correlations induced by the adjacent electrodes, these hybridized states become Andreev molecular states extending across both junctions. The degree of wavefunction delocalization is therefore expected to be maximal at the avoided crossings.

To probe the relation between hybridization and nonlocal Josephson transport, we measure the flux dependence of the Josephson response throughout the stability diagram. Representative traces are shown in Fig.~\ref{fig3}d. A periodic modulation with the same flux periodicity as observed in device~B is recovered over a broad range of gate configurations, demonstrating a robust and reproducible nonlocal Josephson effect in device~T. The strength of this response can be quantified by defining a conductance contrast~$C_G$ which is directly analogous to the switching current contrast introduced for device~B (equation~\ref{eqn:contrast}). The resulting contrast map is shown in Fig.~\ref{fig3}c. We observe that the strongest nonlocal response does not coincide with the conductance resonances defining the stability diagram, but instead that the modulation contrast reaches its maximum at the avoided crossings, where the Andreev molecular wavefunction is expected to be most strongly delocalized. This behavior is observed consistently throughout the stability diagram and becomes particularly pronounced in regions where the avoided crossings are largest, corresponding to stronger interdot hybridization. The systematic enhancement of the nonlocal response at the avoided crossings establishes thus a direct connection between the nonlocal Josephson effect and the formation of delocalized Andreev molecular states.

\section{$\Pi$-phase shift as a probe of global parity}

\begin{figure*}[t]
    \centering
    \includegraphics[width=\textwidth]{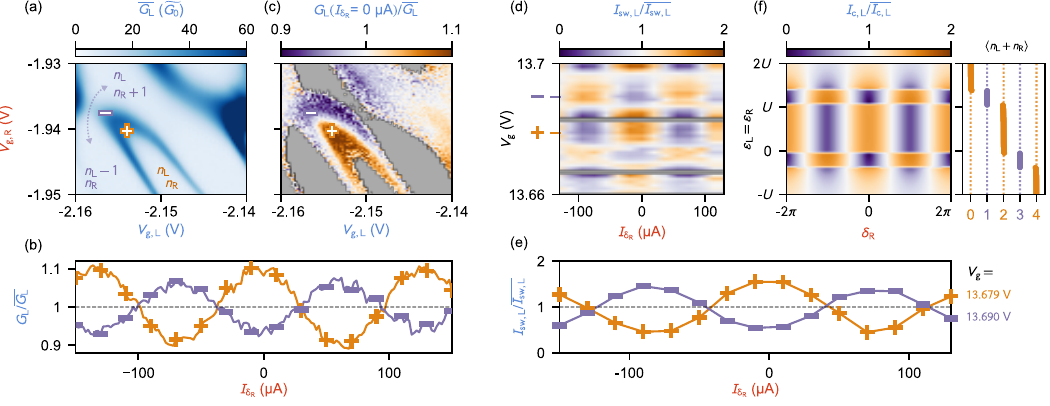}
    \caption{{\bf Probing the global fermion parity through the nonlocal Josephson effect.} (a)~Zero-bias conductance $\overline{G_\mathrm{L}}$ of device~T averaged over~$I_\mathrm{\delta_R}$, as a function of the local gate voltages $V_\mathrm{g,L/R}$. The highlighted region exhibits superconducting avoided crossings connecting charge states differing by two electrons, specified by~$(n_\mathrm{L},n_\mathrm{R})$. (b)~Flux-dependent nonlocal response measured at two operating points of opposite parity (indicated by $+/-$ in (a) and~(c)). (c)~Zero-bias conductance~$G_\mathrm{L}$, normalized by its average value~$\overline{G_\mathrm{L}}$ and measured at fixed flux current~$I_\mathrm{\delta_R}=\SI{0}{\micro\ampere}$, in the same gate-voltage region as in~(a). Abrupt changes in the phase of the nonlocal response, for example from operating points $+$~to~$-$, occur across parity boundaries. Grey regions correspond to gate configurations where the flux dependence was not measured. (d)~Switching current~$I_\mathrm{sw,L}$ of device~B, normalized by its average value $\overline{I_\mathrm{sw,L}}$ over $I_\mathrm{\delta_R}$, as a function of~$V_\mathrm{g}$ and~$I_\mathrm{\delta_R}$. Grey lines correspond to gate voltages where the flux dependence was not measured. (e)~Flux-dependent modulation of the switching current measured at two gate voltages indicated by $+/-$ in~(d). The nonlocal response exhibits opposite phases attributed to changes of global parity. (f)~Theoretical description of the nonlocal response of an Andreev molecule. Left: calculated critical current $I_\mathrm{c,L}$, normalized by its average value $\overline{I_\mathrm{c,L}}$, as a function of~$\delta_\mathrm{R}$ and the common chemical potential $\epsilon_\mathrm{L/R}$ of the two quantum dots scaled by their charging energy~$U$. Right: corresponding ground-state charge. Changes in the phase of the calculated nonlocal response coincide with changes in the ground-state parity.}
    \label{fig4}
\end{figure*}

We now investigate how the parity of this state affects this Josephson response. To this end, we focus on a different region of the stability diagram in device~T, where superconducting correlations are more pronounced and give rise to a well-defined parity structure of the molecular ground state.

The corresponding conductance map is shown in Fig.~\ref{fig4}a. Contrary to the stability diagram discussed previously, charge degeneracy lines now develop avoided crossings that connect charge states differing by two electrons ${(n_\mathrm{L}-1,n_\mathrm{R})\longleftrightarrow(n_\mathrm{L},n_\mathrm{R}+1)}$. Such features reflect the presence of superconducting correlations induced in the quantum dots, which hybridize charge states of identical fermion parity (see Supplementary~\ref*{SM:weak_strong_coupling}). Figure~\ref{fig4}b then shows representative flux-dependent modulations measured at two operating points on opposite sides of these parity boundaries. A clear $\pi$-phase shift is observed: the conductance is maximized at zero flux for one parity configuration and minimized for the other. This behavior is observed throughout the stability diagram, as illustrated in Fig.~\ref{fig4}c, where parity transitions correlate with abrupt shifts in the phase of the nonlocal response.

While device~T provides complete control over the individual charge states, device~B allows one to control the global parity with a single knob,~$V_\mathrm{g}$. On top of that it provides a direct measurement of the switching current itself. The same phenomenology is in fact observed in this second device. As shown in Fig.~\ref{fig4}d-e, the nonlocal modulation of the switching current undergoes abrupt {$\pi$-phase} shifts as the gate voltage is varied. Depending on the gate configuration, the switching current is either maximized or minimized at zero flux, reproducing the behavior observed in device~T and attributed to global parity change. To support this interpretation, we compare the measurements to a minimal model incorporating induced superconducting correlations, interdot hybridization, and Coulomb interactions in a double quantum dot~\cite{meng_self-consistent_2009,feinberg_quartets_2015} (see Methods). Figure~\ref{fig4}f shows the simulated critical current of the left junction~$I_\mathrm{c,L}$ as a function of the phase difference across the right junction~$\delta_\mathrm{R}$ and the common chemical potential of the two dots~$\epsilon_\mathrm{L/R}$. This theoretical calculation reproduces the sequence of {$\pi$-phase} shifts observed experimentally. Moreover, the evaluation of the ground-state charge~${\langle n_\mathrm{L}+n_\mathrm{R}\rangle}$ confirms that each phase shift indeed coincides with a change of the total charge by one electron and therefore with a change of fermion parity. The observation of parity-dependent phase shifts in both conductance and switching-current measurements demonstrates that the effect is robust and independent of the particular detection scheme.

Taken together, these observations demonstrate that the phase of the nonlocal Josephson response directly probes the parity of a delocalized Andreev molecular state. Combined with the results of the previous section, they establish the NLJE as a transport probe that is simultaneously sensitive to the two defining properties of the Andreev molecule: the spatial delocalization of its wavefunction and the fermion parity of its ground state.

\section{Concluding remarks}

In summary, we have demonstrated that the nonlocal Josephson effect provides direct access to both the spatial structure and the fermion parity of a delocalized superconducting state. To do so, we have implemented an elementary Andreev molecule formed by two coupled quantum-dot Josephson junctions using an individual carbon nanotube. Our measurements show that the nonlocal Josephson response reaches its maximum when the molecular wavefunction is delocalized across the coupled junctions, demonstrating the key role of hybridization. Notably, we find that changes in the ground-state parity are encoded as characteristic {$\pi$-phase} shifts in the nonlocal response. This behavior, which mirrors the {$0$--$\pi$} transition of a single quantum-dot Josephson junction, provides a direct transport signature of global fermion parity in an Andreev molecule.

The ability to detect and control global fermion parity of delocalized superconducting states is expected to become increasingly important as quantum coherence is extended across larger superconducting networks. In this context, quantum dot-based Andreev molecules provide a minimal building block for investigating how parity evolves from a local property into a delocalized degree of freedom distributed across coupled junctions. Extending this approach to larger networks could enable the engineering and exploration of topological superconducting phases, including Kitaev-like chains in which quantum information is encoded in delocalized fermionic states. Beyond transport measurements, integrating these systems into circuit-QED architectures could provide high-frequency, quantum non-demolition readout and coherent manipulation of global fermion parity.

\section*{Methods}

\subsection{Device fabrication and measurement setup}
\label{Methods:fabrication_measurement}

The devices were fabricated following the same method as the one detailed in Ref.~\cite{annabi_josephson_2024}. The central electrode is designed to be \SI{100}{\nano\meter} wide. For device~T, a UV-ozone treatment was applied to the niobium/gold electrodes prior to the carbon nanotube transfer in order to improve the electrical contacts between the two. The top gates were patterned using e-beam lithography on PMMA A6 resist and developed in a cold IPA:\ce{H2O} (3:1) mixture. The mask is subsequently metallized by e-beam evaporation, using a sticking layer of \SI{8}{\nano\meter} of titanium covered by \SI{100}{\nano\meter} of aluminium.
The measurement setup is the same as in Ref.~\cite{annabi_josephson_2024}. For device~T, the additional top gate electrodes are connected to \textit{Thermocoax} lines with the same additional infrared and microwave filtering as for the backgate.

\subsection{Flux quantization and phase biasing}
\label{Methods:FluxBias}

The superconducting phase difference across the right junction is controlled by embedding the junction in a superconducting loop threaded by a magnetic flux. Effective phase biasing requires the geometric inductance of the loop, $\mathcal{L}_\mathrm{g}$, to remain negligible compared to the Josephson inductance of the junction, $\mathcal{L}_\mathrm{J}$.
From the device geometry, the loop inductance is estimated to be ${\mathcal{L}_\mathrm{g}\approx\SI{120}{\pico\henry}}$. Near zero phase bias, the Josephson inductance is given by
\begin{equation}
\mathcal{L}_\mathrm{J}=\frac{\varphi_0}{I_0},
\end{equation}
where $\varphi_0=\hbar/2e$ is the reduced flux quantum and $I_0$ is the critical current of the junction. Taking $I_0=\SI{10}{\nano\ampere}$, corresponding to the largest supercurrents observed in our devices~\cite{annabi_josephson_2024}, yields $\mathcal{L}_\mathrm{J}=\SI{30}{\nano\henry}$, more than two orders of magnitude larger than $\mathcal{L}_\mathrm{g}$. The self-inductance of the loop can therefore be neglected, ensuring that the superconducting phase difference across the right junction is directly controlled by the magnetic flux threading the loop.

For both devices, the superconducting loop has dimensions $\SI{9.5}{\micro\meter}\times\SI{65}{\micro\meter}$. The magnetic flux is generated by a current $I_{\delta_\mathrm{R}}$ flowing through a flux line located approximately $\SI{2}{\micro\meter}$ from the loop. Experimentally, the nonlocal Josephson response exhibits a periodic modulation with a flux-line current period of
$\Delta I_{\delta_\mathrm{R}}=\SI{150\pm10}{\micro\ampere}$ for both devices.
Using the loop geometry and the magnetic field generated by the flux line, the corresponding flux variation is
\begin{equation}
\Delta\Phi_\mathrm{R}
=
\iint_{\mathrm{loop}}
\frac{\mu_0 \Delta I_{\delta_\mathrm{R}}}{2\pi y}
\,dx\,dy
=
(2.0\pm0.1)\times10^{-15}\,\mathrm{Wb}.
\end{equation}
This value agrees with the superconducting flux quantum
$\Phi_0=h/2e\approx2.07\times10^{-15}\,\mathrm{Wb}$,
demonstrating that each period of the measured modulation corresponds to the addition of one flux quantum in the loop.
The agreement between the measured periodicity and $\Phi_0$ confirms that the flux-line current provides direct control over the superconducting phase difference across the right junction. However, this geometric calibration does not determine the absolute phase offset of the junction, so that the correspondence between $I_{\delta_\mathrm{R}}$ and $\delta_\mathrm{R}$ remains defined up to an additive constant.

\subsection{Modeling of the Andreev molecule}
\label{Methods:Modeling_Andreev_molecule}

To capture the essential ingredients responsible for the nonlocal Josephson effect and the parity-dependent phase shifts observed experimentally, we use a minimal effective model of two interacting quantum dots. The purpose of the model is not to provide a quantitative description of the device, but rather to identify the physical mechanisms underlying the observed phenomenology. We therefore retain
only the minimal set of ingredients required to reproduce molecular hybridization, nonlocal Josephson coupling, and parity-dependent phase shifts.

\subsubsection{Hamiltonian}
The Hamiltonian reads
\begin{equation}
    \hat{H}=\hat{H}_\mathrm{DQD}+\hat{H}_\mathrm{T}+\hat{H}_\mathrm{S}
\end{equation}
where $\hat{H}_\mathrm{DQD}$ is the Hamiltonian of two isolated and independent quantum dots and writes
\begin{equation*}
    \hat{H}_\mathrm{DQD}=\sum_\mathrm{\alpha=L,R}\sum_{\sigma=\uparrow,\downarrow}
\epsilon_\alpha d^\dagger_{\alpha\sigma} d_{\alpha\sigma}
+ U_\mathrm{L} n_\mathrm{L\uparrow}n_\mathrm{L\downarrow}
+ U_\mathrm{R} n_\mathrm{R\uparrow}n_\mathrm{R\downarrow},
\end{equation*}
the second term is a tunneling Hamiltonian between the two dots
\begin{equation*}
    \hat{H}_\mathrm{T}=t \sum_{\sigma=\uparrow,\downarrow}
\left[
e^{i(\delta_\mathrm{R}-\delta_\mathrm{L})/4}
d^\dagger_\mathrm{L\sigma} d_\mathrm{R\sigma}
+
\mathrm{h.c.}
\right],
\end{equation*}
and the last term is the pairing induced in the quantum dots by the connection to the superconducting electrodes
\begin{equation*}
    \hat{H}_\mathrm{S}=\sum_\mathrm{\alpha=L,R}
\left(
\Gamma_\alpha \cos\frac{\delta_\alpha}{2}
+i\delta\Gamma_\alpha \sin\frac{\delta_\alpha}{2}
\right)
d^\dagger_{\alpha\uparrow}d^\dagger_{\alpha\downarrow}
+\mathrm{h.c.}
\end{equation*}

Here, $d^\dagger_{\alpha\sigma}$ creates an electron with spin $\sigma$ on dot $\alpha=\mathrm{L,R}$, $n_{\alpha\sigma}=d^\dagger_{\alpha\sigma}d_{\alpha\sigma}$, $\epsilon_{\alpha}$ are the dot orbital energies, $U_{\alpha}$ are the charging energies, and $t$ is the interdot tunnel coupling. The phases $\delta_\mathrm{L}$ and $\delta_\mathrm{R}$ are the superconducting phase differences across the left and right junctions. The parameters $\Gamma_\mathrm{L}$ and $\Gamma_\mathrm{R}$ describe the proximity-induced local pairing on each dot, while $\delta\Gamma_\mathrm{L}$ and $\delta\Gamma_\mathrm{R}$ account for left-right asymmetry of the superconducting couplings within each junction.

To retain a minimal description and minimize the number of free parameters, we neglect both the interdot Coulomb interaction and nonlocal pairing terms associated with crossed Andreev processes. More specifically, an interdot interaction term $U_\mathrm{LR}n_\mathrm{L}n_\mathrm{R}$ and nonlocal pairing terms of the form $\Gamma_\mathrm{LR}d^\dagger_\mathrm{L\uparrow}d^\dagger_\mathrm{R\downarrow}$ can improve the quantitative agreement with some features of the experimental stability diagrams. However, neither is required to reproduce the central observations reported here, namely molecular hybridization, nonlocal Josephson coupling, and parity-dependent phase shifts. All calculations presented in this work were therefore performed with $U_\mathrm{LR}=0$ and without nonlocal pairing terms, i.e. $\Gamma_\mathrm{LR}=0$.

\subsubsection{Critical current and charge state}
For a given set of parameters, the Hamiltonian is diagonalized exactly in the full many-body Hilbert space. The ground-state energy $E_0(\delta_\mathrm{L},\delta_\mathrm{R})$ is then used to compute the Josephson current through the left junction,
\begin{equation}
I_\mathrm{L}(\delta_\mathrm{L},\delta_\mathrm{R})
= \frac{2e}{\hbar}\frac{\partial E_0}{\partial \delta_\mathrm{L}}.
\end{equation}
The critical current plotted in Fig.~4f of the main text is obtained by maximizing this current over the phase difference $\delta_\mathrm{L}$,
\begin{equation}
I_\mathrm{c}(\delta_\mathrm{R})=\max_{\delta_\mathrm{L}} I_\mathrm{L}(\delta_\mathrm{L},\delta_\mathrm{R}).
\end{equation}

In Fig.~\ref{fig4}f of the main text, the two dot energies are swept together, $\epsilon_\mathrm{L}=\epsilon_\mathrm{R}=\epsilon$, mimicking the action of the global back-gate in device~B, and the parameters are $\Gamma_\mathrm{L}=\Gamma_\mathrm{R}=0.5$, $\delta\Gamma_\mathrm{L,R}=0.25$, $t=2$ and $U_\mathrm{L}=U_\mathrm{R}=5$. The calculated critical current exhibits periodic oscillations with $\delta_\mathrm{R}$, whose phase changes abruptly as $\epsilon$ is varied. If we define $|\Psi_0\rangle$ as the many-body ground state, its corresponding charge
\begin{equation}
N=\sum_\mathrm{\alpha=L,R}\sum_{\sigma=\uparrow,\downarrow}
\langle \Psi_0|n_{\alpha\sigma}|\Psi_0\rangle
\end{equation}
changes by one electron at the same values of $\epsilon$, showing that the phase shifts of the nonlocal Josephson response coincide with changes of the global fermion parity of the molecular ground state.

\begin{acknowledgments}
We acknowledge valuable discussions with H. Duprez, F. Nichele, M. Houzet, Ç. Girit, A. Keliri, A. Peugeot, R. Deblock, P. Joyez and M. Delbecq. Special thanks go to R. Mohammedi, D. Roux and T. Pépin-Donat for their technical support. Gratitude is extended to R. Ribeiro-Palau, P.F. Orfila, S. Delprat and the CEA Saclay Nanofab team for their help on nanofabrication processes. 
\end{acknowledgments}

\section*{Fundings}

L.B. acknowledges support of the European Research Council (ERC) under the European Union’s Horizon 2020 research and innovation programme (Grant Agreement No. 947707). 
J.-D.P. acknowledges support from the Agence Nationale de la Recherche (Grant No. ANR-20-CE47-0003).
L.B. and J.-D.P. acknowledge the support of the Fondation de l'École polytechnique.
This work has been supported by the French ANR-22-PETQ-0003 Grant under the France 2030 plan. 
K.W. and T.T. acknowledge support from the JSPS KAKENHI (Grant Numbers 21H05233 and 23H02052), the CREST (JPMJCR24A5), JST, and World Premier International Research Center Initiative (WPI), MEXT, Japan.

\clearpage

\renewcommand{\thefigure}{S\arabic{figure}}
\setcounter{figure}{0}
\setcounter{section}{0}

\onecolumngrid

\begin{center}
    {\LARGE\bfseries Supplementary Information for\\
    Fermion parity of an Andreev molecule probed by nonlocal Josephson effect\par}
    {\large }
\end{center}

\section{Ballistic transport in device~B}
\label{SM:Ballistic}

\begin{figure}[h]
    \centering
    \includegraphics[width=\columnwidth]{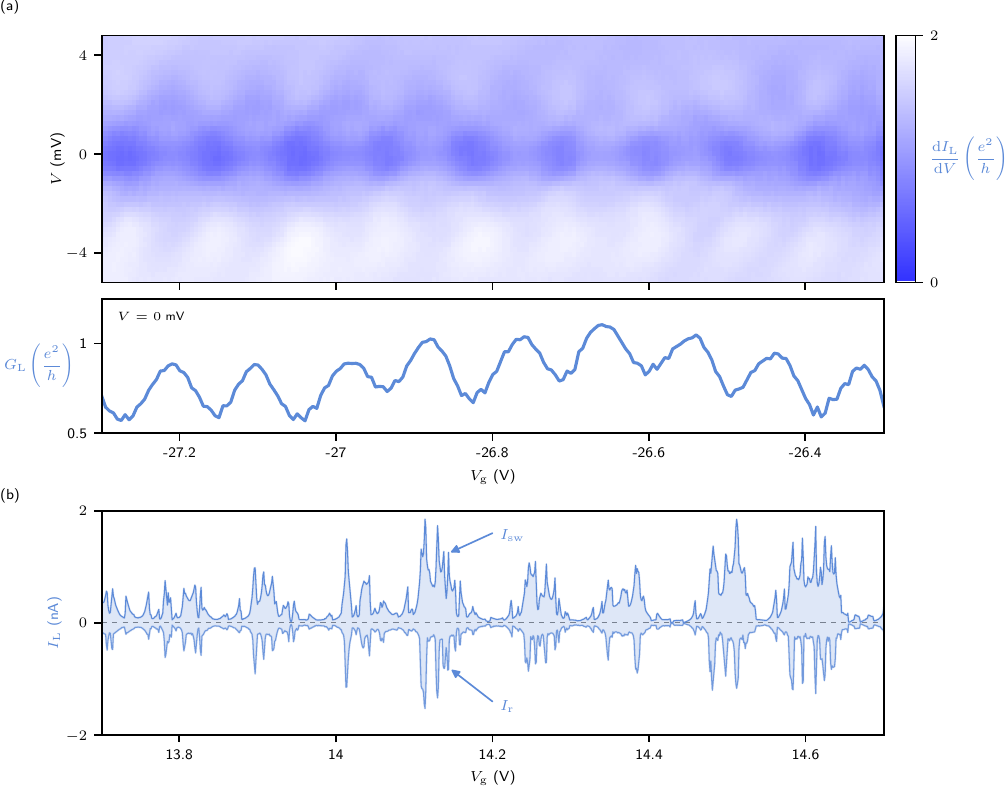}
    \caption{{\bf Ballistic transport in device~B.} (a) Differential conductance measured at \SI{3.5}{\kelvin}, above the superconducting critical temperature of the electrodes, as a function of gate voltage $V_\mathrm{g}$ and bias voltage $V$. The checkerboard pattern and the corresponding oscillations of the zero-bias conductance (bottom panel) are characteristic signatures of Fabry-Pérot interference in a ballistic nanotube cavity. (b) Switching current $I_\mathrm{sw}$ and retrapping current $I_\mathrm{r}$ of device~B measured at \SI{10}{\milli\kelvin} as a function of gate voltage. The broad modulation of the supercurrent amplitude follows the same periodicity as the Fabry--Pérot oscillations observed in the normal state, indicating that the Josephson transport is governed by the same ballistic electronic states.}
    \label{figSM:fabryperot}
\end{figure}

When the electrodes of device~B are driven into the normal state by measuring above their critical temperature, the differential conductance exhibits a characteristic checkerboard pattern together with regular conductance oscillations as a function of gate voltage (Fig.~\ref{figSM:fabryperot}a). These features are signatures of Fabry-Pérot interference in the nanotube cavity formed between the electrodes and indicate ballistic electronic transport.
The characteristic energy scale of the interferometer can be extracted from the checkerboard pattern~\cite{liang2001fabry}. Identifying this energy with the confinement energy ${\Delta E = hv_\mathrm{F}/2L}$, where $v_\mathrm{F}$ is the Fermi velocity and $L$ the cavity length, yields $L\approx\SI{700}{\nano\meter}$ for ${\Delta E\approx\SI{2.5}{\milli\eV}}$. This value is in good agreement with the lithographic separation between the superconducting contacts.

At low temperature, the switching current of the left junction exhibits a broad modulation with gate voltage that follows the same periodicity as the Fabry-Pérot oscillations observed in the normal state (Fig.~\ref{figSM:fabryperot}b). This correspondence suggests that the Josephson transport is governed by the same ballistic electronic states responsible for the normal-state interference pattern.

\clearpage

\section{Zero-bias conductance as a probe of the Josephson response}
\label{SM:Conductance_Josephson}

\begin{figure}[h]
    \centering
    \includegraphics[width=\columnwidth]{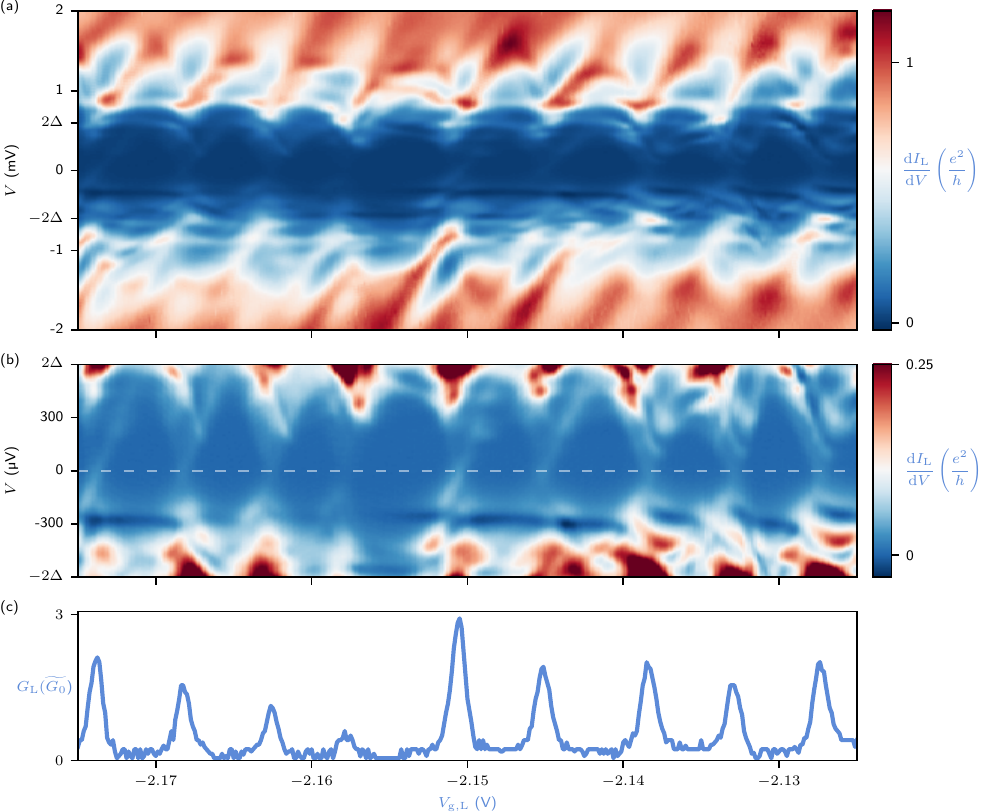}
    \caption{{\bf Zero-bias conductance as a probe of the Josephson response in device~T.} (a) Differential conductance measured at \SI{10}{\milli\kelvin} as a function of the left gate voltage $V_\mathrm{g,L}$ and bias voltage $V$. The low-conductance region around zero bias corresponds to the superconducting gap~${\Delta=\SI{300}{\micro\eV}}$ of the electrodes, estimated from independent measurements~\cite{annabi:tel-05265252}. (b) Close-up of the subgap region showing gate-dependent resonances associated with Andreev transport through the device. (c) Zero-bias conductance extracted from (b). Pronounced conductance peaks appear at the charge degeneracy points of the left quantum dot, where the Josephson coupling is expected to be strongest. Throughout this work, the amplitude of the zero-bias conductance is used as a proxy for the local Josephson response in device~T.}
    \label{figSM:conductance_josephson}
\end{figure}

The transport characteristics of device~T differ from those of device~B. While device~B exhibits a well-defined switching current associated with a visible supercurrent branch, the Josephson response of device~T is primarily revealed through an enhanced zero-bias conductance. Figure~\ref{figSM:conductance_josephson}a shows the differential conductance of device~T as a function of bias voltage and gate voltage. A low-conductance region is observed around zero bias, corresponding to the superconducting gap of the electrodes. At lower bias voltages, additional gate-dependent resonances emerge (Fig.~\ref{figSM:conductance_josephson}b), giving rise to pronounced peaks in the zero-bias conductance (Fig.~\ref{figSM:conductance_josephson}c). 

Several observations support the interpretation of this signal as originating from Josephson transport. First, the zero-bias conductance follows the charge degeneracy lines of the quantum dot and is maximal at the resonances where the supercurrent is expected to be largest. Most importantly, its amplitude exhibits a pronounced periodic modulation with magnetic flux, with a period corresponding to one superconducting flux quantum $h/2e$. This modulation is observed only in the regime of strong interdot hybridization. This demonstrates that the measured signal originates from phase-coherent superconducting transport through the loop and therefore from the Josephson effect.

A natural interpretation of these observations is that device~T operates in a regime of phase diffusion. In Josephson junctions with relatively small critical currents, thermal activation and electrical noise induce phase fluctuations that transform the ideal dissipationless supercurrent branch into a finite-slope current-voltage characteristic~\cite{martinis_classical_1989,vion_thermal_1996}. In the strong phase-diffusion regime, the resulting feature appears as an enhanced differential conductance centered at zero voltage rather than as a sharply resolved switching current. Such behavior has previously been reported in carbon nanotube Josephson junctions with critical currents of only a few nanoamperes~\cite{jorgensen_critical_2007,eichler_tuning_2009}. Similar transport characteristics have also been reported in the most advanced realizations of double quantum-dot Josephson junctions, including InAs and InSb-based S-DQD-S devices~\cite{estrada_saldana_supercurrent_2018,su_andreev_2017,bouman_triplet-blockaded_2020,prosko_flux-tunable_2024}, where the Josephson response is typically revealed through low-bias transport signatures rather than through a sharply defined dissipationless branch.

Several factors may contribute to the stronger phase diffusion observed in device~T compared to device~B. Device~T incorporates two local top gates in close proximity to the junctions, in addition to the back gate, thereby increasing the number of potentially noisy electrical lines coupled to the device. Furthermore, its normal-state resistance is approximately twice larger than that of device~B, implying a smaller critical current~\cite{annabi_josephson_2024} and therefore a greater sensitivity to thermal and electrical fluctuations.

Throughout the main text, the amplitude of this conductance is therefore used as a proxy for the local Josephson response and its nonlocal modulation, complementary to the switching-current measurements performed on device~B.

\clearpage

\section{Gate dependence of the nonlocal Josephson contrast}
\label{SM:Isw_CI_vs_gate}

\begin{figure}[h]
    \centering
    \includegraphics[width=\columnwidth]{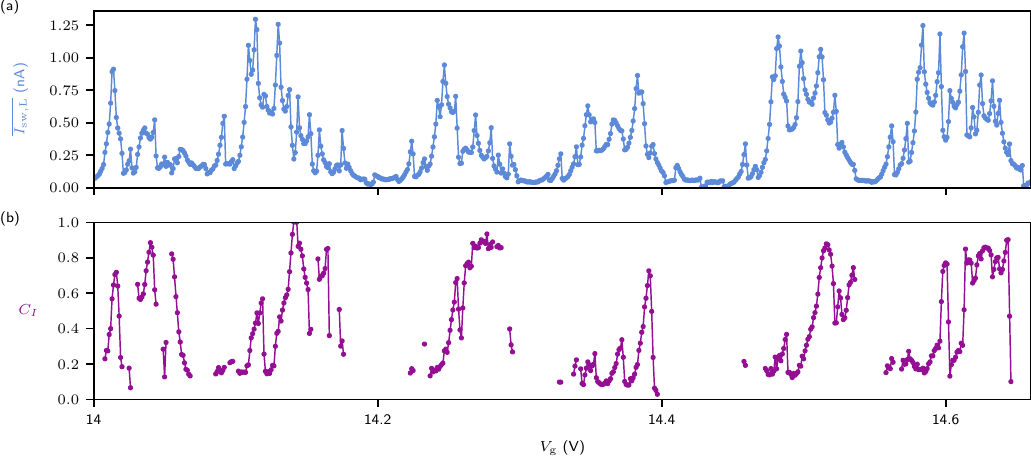}
    \caption{{\bf Gate dependence of the nonlocal Josephson contrast in device~B.} (a) Average switching current as a function of gate voltage. The average switching current is defined as the mean value of the maximum and minimum switching currents measured over one flux period. (b) Corresponding nonlocal Josephson contrast extracted from the modulation of the switching current by the phase difference across the right junction. The contrast can be followed over a broad gate-voltage range and reaches values close to unity.}
    \label{figSM:Contrast}
\end{figure}

Figure~\ref{figSM:Contrast} shows the evolution of the average switching current and of the nonlocal Josephson contrast as a function of gate voltage in device~B. The average switching current is defined as the mean value of the maximum and minimum switching currents measured over one flux period.

The switching current exhibits broad gate-dependent variations, reflecting the evolution of the Josephson coupling as the electronic states of the nanotube are tuned. The corresponding nonlocal modulation contrast can be followed over an extended gate voltage range and varies relatively smoothly between nearly zero and values approaching unity, demonstrating that the nonlocal Josephson effect can be efficiently controlled by electrostatic gating. Importantly, the contrast does not exhibit a simple correlation with the magnitude of the switching current, suggesting that the strength of the nonlocal response is governed by the microscopic properties of the Andreev molecular state rather than solely by the Josephson coupling of the measured junction.

\clearpage

\section{Switching current measurements}
\label{SM:Switching_measurement}

Throughout the main text, the switching current $I_\mathrm{sw,L}$ is extracted directly from the current-voltage characteristics as the largest current belonging to the zero-voltage branch. This procedure enables the rapid acquisition of the large datasets required to map the nonlocal Josephson response as a function of gate voltage and magnetic flux. 

\begin{figure}[h]
    \centering
    \includegraphics[width=\columnwidth]{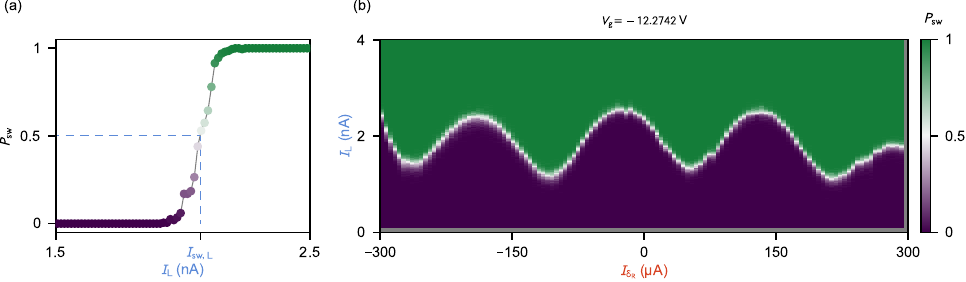}
    \caption{{\bf Switching-current measurements.} (a) Switching probability $P_\mathrm{sw}$ of device~B as a function of the current-pulse amplitude $I_\mathrm{L}$, measured at fixed gate voltage and flux. The characteristic sigmoidal dependence expected for stochastic switching in a Josephson junction is observed. The switching current $I_\mathrm{sw,L}$ is defined as the current corresponding to $P_\mathrm{sw}=0.5$. (b) Switching probability $P_\mathrm{sw}$ as a function of the pulse amplitude $I_\mathrm{L}$ and flux-line current $I_{\delta_\mathrm{R}}$ at a representative gate voltage~$V_\mathrm{g}$. The position of the switching transition (white line) defines the switching current $I_\mathrm{sw,L}$ and exhibits a clear periodic modulation with magnetic flux, revealing the nonlocal Josephson effect. The amplitude and periodicity of this modulation are consistent with those obtained from the direct analysis of the current-voltage characteristics used throughout the main text. For these measurements, 200 square pulses with \SI{200}{\micro\second} pulse width are applied at a repetition rate of \SI{312.013}{\hertz}.}
    \label{figSM:Psw}
    \end{figure}

To validate this approach, we additionally performed conventional switching-probability measurements~\cite{della_rocca_measurement_2007} on device~B. Figure~\ref{figSM:Psw}a shows a typical switching curve obtained by applying repeated current pulses of fixed amplitude~$I_\mathrm{L}$ to the junction. The resulting sigmoidal dependence of the switching probability ${P_\mathrm{sw}(I_\mathrm{L})}$ on the pulse amplitude is characteristic of stochastic switching in a Josephson junction. The switching current is defined as the current corresponding to ${P_\mathrm{sw}(I_\mathrm{sw,L})=0.5}$.

The evolution of the switching probability as a function of the flux current $I_{\delta_\mathrm{R}}$ is shown in Fig.~\ref{figSM:Psw}b at a representative gate voltage. The position of the switching transition is periodically modulated by the superconducting phase difference across the right junction, directly revealing the nonlocal Josephson effect. The width of the switching transition is significantly smaller than the amplitude of the modulation, allowing the latter to be resolved with high precision.

The modulation contrast extracted from these switching measurements is fully consistent with that obtained from the direct analysis of the current-voltage characteristics used throughout the main text. These measurements therefore validate the simpler procedure employed in the main text to determine the switching current and confirm the robustness of the observed nonlocal Josephson response.

\clearpage

\section{Extended double quantum-dot stability diagram}
\label{SM:DQD_stability_wide}

\begin{figure}[h]
    \centering
    \includegraphics[width=\columnwidth]{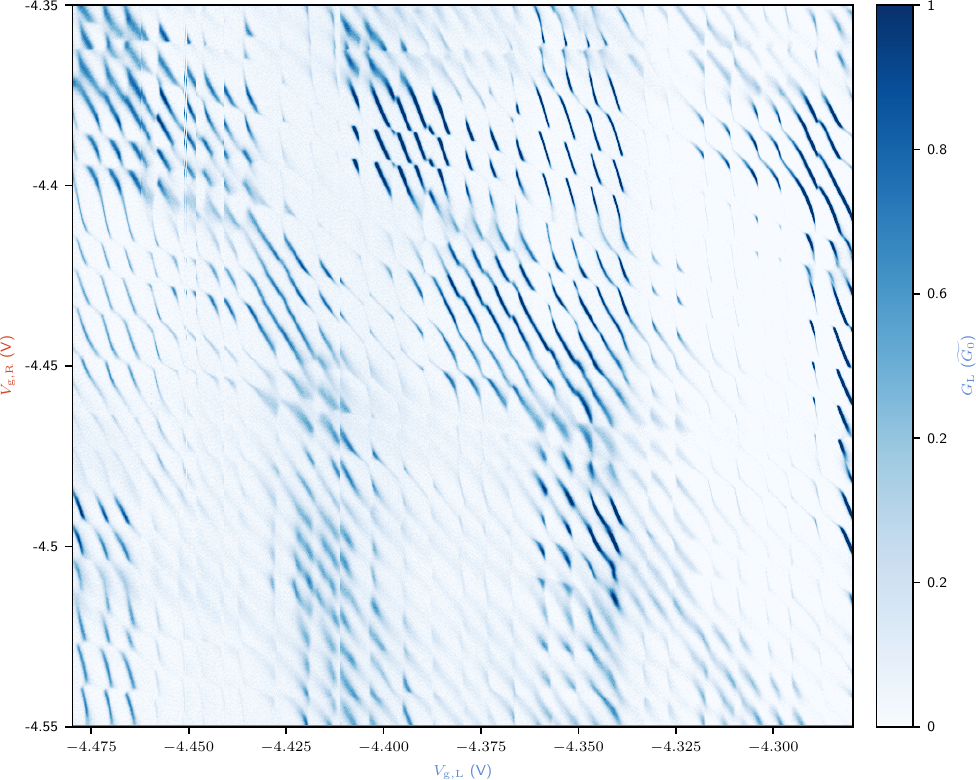}
    \caption{{\bf Extended stability diagram of device~T.} Zero-bias conductance~$G_\mathrm{L}$ measured as a function of the two local gate voltages over a broad range of electron occupancies. Different regions exhibit qualitatively distinct charge-stability patterns, ranging from weakly coupled double-dot behavior to strongly hybridized molecular states.}
    \label{figSM:Stability}
\end{figure}

Figure~\ref{figSM:Stability} shows a conductance map of device~T measured over a broader range of gate voltages than the data presented in the main text. The stability diagram reveals a rich evolution of the electronic states of the double quantum-dot system as the occupation is varied.

Depending on gate voltages, the device exhibits markedly different charge-stability patterns. In some regions, the two quantum dots appear weakly coupled and retain distinct charge degeneracy lines, while in others the charge states become strongly hybridized and the stability diagram approaches that of a single extended quantum system. Intermediate regimes are also observed, characterized by avoided crossings with varying degrees of curvature and sharpness, reflecting different relative contributions of tunnel coupling and electrostatic interactions.

These measurements highlight the diversity of electronic regimes accessible in the same device and demonstrate that the coupling between the two quantum dots can vary substantially with gate voltage. The data also illustrates the complexity of the underlying double-dot spectrum and provide broader context for the gate-voltage region investigated in the main text, where the nonlocal Josephson effect is characterized in detail.

\clearpage

\section{Weak and strong-induced pairing regimes}
\label{SM:weak_strong_coupling}

The model is also used to illustrate the two regimes observed experimentally in devices~T and~B. We compute the ground-state properties as a function of the two independent dot chemical potentials $\epsilon_\mathrm{L}$ and $\epsilon_\mathrm{R}$, which correspond experimentally to the two local gate voltages of device~T. In both cases, the calculated quantity is the critical current through the left junction, demonstrating that local transport measurements can reveal global properties of the Andreev molecule.

\begin{figure}[h]
    \centering
    \includegraphics[width=\columnwidth]{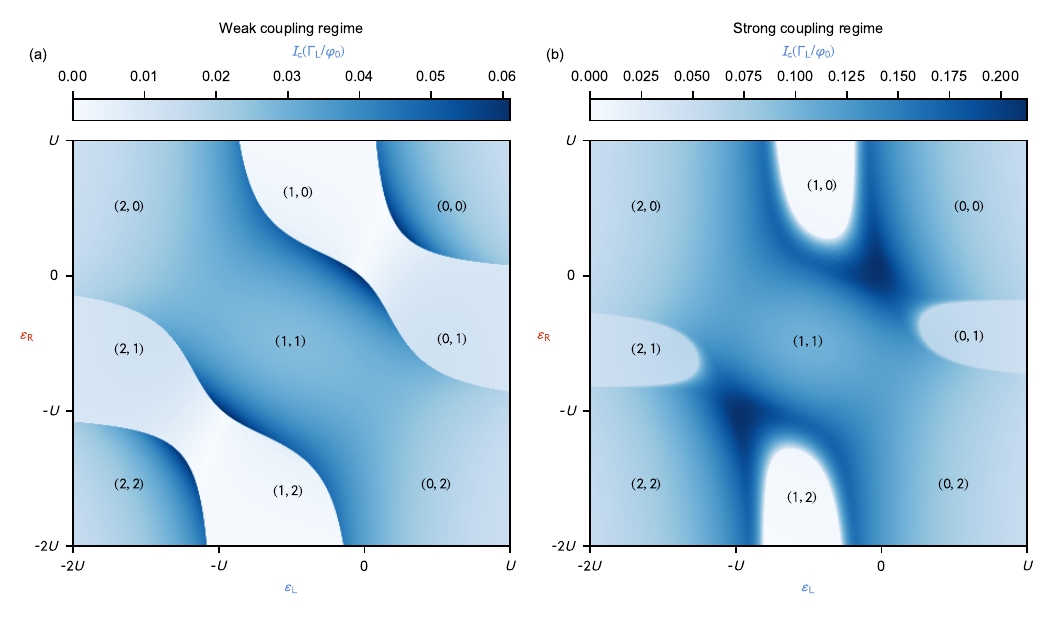}
    \caption{{\bf Calculated stability diagrams using the minimal Andreev-molecule model.} (a)~Critical current $I_\mathrm{c}$ of the left junction as a function of the two dot energies $\epsilon_\mathrm{L,R}$, in the weak induced-pairing regime. The parameters are $U_\mathrm{L,R}=10$, $t=3$, $\Gamma_\mathrm{L,R}=1$, $\delta\Gamma_\mathrm{L,R}=0.5$, and $\delta_\mathrm{R}=0$. Two families of charge degeneracy lines associated with the left and right dots are visible. Their intersections develop avoided crossings due to interdot hybridization, signaling the formation of delocalized molecular states. The resulting stability diagram closely resembles that of a conventional double quantum dot and qualitatively reproduces the regime observed experimentally in Fig.~3a of the main text. The labels $(n_\mathrm{L},n_\mathrm{R})$ indicate the dominant charge configuration of the ground state, where $n_\mathrm{L,R}$ denote the occupations of the left and right dots, respectively. (b)~$I_\mathrm{c}$ as a function of $\epsilon_\mathrm{L,R}$, in the strong induced-pairing regime. The parameters are $U_\mathrm{L,R}=5$, $t=1$, $\Gamma_\mathrm{L,R}=2$, $\delta\Gamma_\mathrm{L,R}=1$, and $\delta_\mathrm{R}=0$. Strong induced superconducting correlations hybridize charge states with identical fermion parity, generating avoided crossings that surround characteristic parity islands. As a consequence, the ground state can evolve continuously between configurations differing by two electrons, corresponding to the addition or removal of a Cooper pair, without undergoing an abrupt parity transition. This regime qualitatively reproduces the stability diagram observed experimentally in Fig.~4a of the main text and highlights the superconducting origin of the parity structure. Owing to superconducting pairing, charge states with identical fermion parity become hybridized and are no longer separated by sharp charge degeneracy lines. In particular, even-parity configurations $(2p,2q)$ become continuously connected through the parity islands visible in the stability diagram, allowing the ground state to evolve smoothly between occupations differing by a Cooper pair.}
    \label{figSM:Simu_weak_strong_coupling}
\end{figure}

For weak induced superconducting correlations, the calculated stability diagram resembles that of a conventional double quantum dot (Fig.~\ref{figSM:Simu_weak_strong_coupling}a). The charge degeneracy lines associated with the left and right dots form two families of resonances, whose intersections develop avoided crossings due to interdot hybridization. This regime corresponds to the experimental situation of Fig.~3a of the main text: the double-dot spectrum is well resolved, and the nonlocal Josephson response is strongest at the avoided crossings where the molecular wavefunction is maximally delocalized.

For stronger induced superconducting correlations (Fig.~\ref{figSM:Simu_weak_strong_coupling}b), the stability diagram changes qualitatively. Superconducting pairing hybridizes charge states with the same fermion parity, producing avoided crossings that connect configurations differing by two electrons. As a result, neighboring even-parity charge states become continuously connected, forming extended parity regions within which the ground state evolves smoothly between charge configurations differing by a Cooper pair. This regime reproduces the qualitative structure observed experimentally in Fig.~4a of the main text, where the stability diagram directly reveals the parity structure of the Andreev molecular ground state.

The parameters used for the representative calculations are chosen to capture the qualitative regimes observed experimentally rather than to provide a quantitative fit. The weak and strong induced-pairing stability diagrams are obtained by varying the induced pairing amplitudes $\Gamma_\mathrm{L,R}$ relative to $t$ and $U$. Increasing the induced superconducting correlations transforms the double-dot stability diagram from a predominantly single-electron structure into a superconducting parity diagram.

\bibliographystyle{apsrev4-2}
\bibliography{NLJE_biblio}

\end{document}